\documentclass[review]{elsarticle}
\usepackage{amssymb}

\journal{Physica A: Statistical Mechanics and its Applications}

\begin{document}

\begin{frontmatter}

\title{Thermodynamics of Equilibrium Alkali Plasma.\\ Simple and Accurate Analytical Model\\ for Non-Trivial Case}

\author[adr1,adr2]{Anatolii V.~Mokshin\corref{cor1}}
\ead{anatolii.mokshin@mail.ru}\cortext[cor1]{Corresponding author}
\author[adr1]{Diana A. Mirziyarova}

\address[adr1]{Department of Computational Physics, Institute of Physics, Kazan
Federal University, 420008 Kazan, Russia}
\address[adr2]{Udmurt Federal Research Center, Ural Branch of Russian Academy of Sciences, 426068 Izhevsk, Russia}

\begin{abstract}
Analytical equation of state for pure alkali metals (lithium, sodium, potassium, rubidium and cesium) in equilibrium gas phase is proposed. This equation has a simple form, generalizes the equation of state for a perfect gas and is universal for all alkali elements. It correctly reproduces the experimental data for the equilibrium gas phase over a wide range of pressures (up to $\sim 10^5$~Pa) and temperatures (up to $\sim 3 \cdot 10^3$~K). On the basis of this equation of state, expressions for thermal and caloric coefficients as well as for \textit{other physical characteristics} are obtained. The results of this study confirm feasibility of the principle of corresponding states in relation to the group of alkali elements.
\end{abstract}

\begin{keyword}
Equations of state, alkali metals, thermodynamic properties, corresponding states, thermodynamic similarity
\end{keyword}

\end{frontmatter}

\section{Introduction}

The equation of state (EoS) is a key element in the thermodynamic description of a physical system, and knowledge of EoS is a \textit{necessary condition} for such a description of the system under consideration. This equation directly determines the thermodynamic conditions under which the equilibrium phases (solid, liquid, gas) appear. Some types of EoS can be used to determine the phase separation conditions (melting, condensation \textit{etc}.) and to predict the thermodynamic properties of the considered system at extremal conditions (for example, at extremely high temperatures and pressures)~\cite{Fortov_Extereme}. In addition, on the basis of this equation for the concrete system, the physical parameters of phase transitions -- such as the free nucleation energy and the driving force of the phase transition -- can be evaluated.

Thermal EoS sets a relation $f(p,\,V,\,T)=0$ between the pressure $p$, the volume $V$ and the temperature $T$ of equilibrium homogeneous thermodynamic system~\cite{Kubo_thermodynamics,Yang_Physa,Wang_Physa}. In this equation, instead of the volume $V$, the parameter can be molar volume $V_m$ or density $\rho$. Since this equation is unique for the concrete system, then it cannot be obtained just from the common thermodynamic relations. Usually, it is determined experimentally. It is very difficult to recover this equation in analytical form even for \textit{an equilibrium liquid or gas phases} without taking into account the phase transitions between both the phases~\cite{Physa_3}. This can be done only for the simplest isotropic model systems for which the microscopic structure and interparticle interactions are unambiguously defined~\cite{Hansen/McDonald} and the virial thermal EoS
\begin{equation} \label{eq: virial_gen}
    \frac{p}{RT} = A_1 \rho + \sum_{i=2} A_i(T) \rho^i
\end{equation}
with the theoretically defined coefficients $A_1\equiv\mu^{-1}$, $A_2(T)$, $A_3(T)$, $\ldots$ is reduced to the form
\begin{equation} \label{eq: virial_approx}
    \frac{p}{RT} =  \rho \frac{1}{\mu} - \rho^2  \frac{2\pi R }{3(k_B\mu)^2 T}  \int_{0}^{\infty}  \frac{du(r)}{dr} \, r^3 \, g(r,p,T)\, dr.
\end{equation}
Here, $R$ is the universal gas constant; $\rho=m/V$ is the mass density; $m$ and $V$ is the mass and volume of the system, respectively; $\mu$ is the molar mass; the radial distribution function $g(r,p,T)$ and the \textit{pairwise additive} potential $u(r)$ specifying particle interactions should be known.

It seems that alkali metals are the simplest metals from the point of view of interparticle (interatomic) interactions, where each atom contains only one valence electron. For these metals, the various effective interatomic pair potentials were developed within the pseudopotential models~\cite{Mokshin/Yulmetyev_JCP,Chekmarev_JCP}: the Ashcroft's model~\cite{Kahl_MD,Alemany_JCP}, the Shaw's model~\cite{Tanaka_MD}, the Price-Singwi-Tosi model~\cite{Price_MD} and the model of Fiolhais \textit{et al.}~\cite{Fiolhais_MD}. These pair potentials have been found to produce results for the microscopic structure properties and the transport coefficients for liquid alkali metals that are in good agreement with experimental data~\cite{Balucani_1992,Wax_Bretonnet,Wax_2019,AVM_JETP_Lett_2017,KhRM/AVM/GBN_JETP_2018,Transport_properties_JCP}. Therefore, one can expect that the thermodynamic properties of liquid alkali metals will be correctly reproduced using virial equation~(\ref{eq: virial_approx}), where the potential $u(r) $ is required as input.  Thermodynamics of alkali metals in the equilibrium gas (vapor) phase differs significantly from the case of the liquid phase due to pronounced ionization effects~\cite{Potential_energy_JCP}. In fact, vapors of alkali metals near their saturation lines are a plasma-like substance, which is a mixture of electron gas and clusters of various charges and sizes~\cite{Fortov_plasma,Physa_5}. The concentration of these clusters-ions is highly dependent on thermodynamic conditions (and, in particular, on temperature). Proper accounting for all the interaction effects can be done using the advanced theories developed in the framework of statistical mechanics (see, for example, Refs.~\cite{Klimontovich,Sadykova,Stone_Rb}). On the other hand, virial equation~(\ref{eq: virial_approx}) can only be used to obtain estimates and approximate thermodynamic results with the effective particular potentials such as the Hellmann potential and the Hellmann–Gurski–Krasko potential~\cite{Reich_Ebeling}. Common virial EoS's for alkali vapors found by fitting Eq.~(\ref{eq: virial_gen}) to the experimental data and given in the handbook~\cite{Ohse_Handbook} include from four to seven contributions in virial expansion~(\ref{eq: virial_gen}) and may contain twenty or more parameters~\cite{Mozg_Rb,Zhukhov,Zakharova}. In such a situation, the following question arises quite naturally: Is it possible to formulate a simple \textit{effective} analytical model that could reproduce the experimental data for thermal EoS of alkali elements in a gas phase?

\section{Thermodynamics. Analytical results}

The equilibrium gas phase of alkali elements had previously been studied experimentally in sufficient detail~\cite{Ohse_Handbook,Stone_data,Stishov_1976,Shpilrain_book,Dillon_JCP}. In particular, experimental correspondence between the pressure~$p$, the temperature~$T$ and the density~$\rho$ of lithium, sodium, potassium, rubidium and cesium in the gas phase is known in tabular form~\cite{Ohse_Handbook,Vargaftik}. These data cover the pressure range from $1$~Pa to $1$~MPa and the temperature range from $600$~K to $3\,000$~K. In the units of the critical pressure $p_c$ and the critical temperature $T_c$ for alkali elements, the considered thermodynamic range corresponds to the pressures
$p\in[10^{-8},\;10^{-1}]$\,$p_c$ and the temperatures $T\in[0.17,\;1.5]$\,$T_c$. Values of the critical point parameters are given in Tab.~\ref{tab: 1}. Thus, the available experimental data for EoS cover the thermodynamic region of an equilibrium gas including the ($p,T$)-states with the temperatures greater than the critical temperature~$T_c$.
\begin{table}[ht]
    \centering
    \caption{Some properties of alkali metals: atomic mass $\mu$ ($10^{-3}$\,kg/mol), critical temperature $T_c$ (K), pressure $p_c$~(MPa) and density $\rho_c$ (kg/m$^3$).}
    \label{tab: 1}
\begin{tabular}{l|ccccr}
    \hline
        & $\mu$   & $T_c$           & $p_c$          & $\rho_{c}$     &                        \\
    \hline \hline
Li      & $6.94$  & $3503$ & $38.42$ & $110.4$ & \cite{Thermophys_prop}  \\
Na      & $22.99$ & $2497$ & $25.22$  & $212$     & \cite{Thermophys_prop}  \\
K       & $39.1$  & $2239$ & $15.95$ & $192$ & \cite{Thermophys_prop}   \\
Rb      & $85.47$ & $2096$  & $13.4$ & $350$ & \cite{Vargaftik}                          \\
Cs      & $132.91$& $2035$  & $11.46$ & $425$ & \cite{Thermophys_prop}  \\
        \hline
        \hline
    \end{tabular}
\end{table}

\subsection{Thermal equation of states}
The experimental data for EoS reveal the following important patterns, which seem to be universal for alkali elements.
First, along the isotherms, correspondence between the data for pressure and density (volume) is well approximated by simple linear relation
\begin{equation} \label{eq: first_property}
    \left( \frac{p}{\rho} \right )_T = \textrm{const},
\end{equation}
that coincides with the Boyle–Mariotte law for a perfect gas.
Second, correlation between the pressure $p$ and the temperature $T$  at isochores differs significantly from the Gay-Lussac's law, and it obeys
\begin{equation} \label{eq: second_property}
    \left ( \frac{p}{T} \right )_\rho \neq \textrm{const}.
\end{equation}
Namely, with an increase in the temperature $T$, the pressure $p$ grows much faster than in the case of a perfect gas.
Third, under isobaric conditions, the density $\rho$ decreases with the increasing temperature much faster than according to the Charles's law. This empirical fact can be generally formulated as
\begin{equation} \label{eq: third_property}
    \left ( \rho T \right )_{p} \neq \textrm{const},
\end{equation}
and it is closely related to the previous two. It is assumed that these patterns are extending up to vapor states with extremely high temperatures~\cite{Li_high_temp_plasma}.

From the analysis of the experimental thermodynamic data for all the considered alkali elements we find that the thermal EoS  for the $(p,T)$-range corresponding to equilibrium gas phase can be defined as
\begin{equation} \label{eq: EoS_PVT}
    pV = k \frac{m}{\mu}\, 10^{b T} \ \ \ {(\mathrm{with}\ \mathrm{the}\ \mathrm{variables}\ p,\ V \ \mathrm{and}\ T)},
\end{equation}
or as
\begin{equation} \label{eq: EoS_PRhoT}
    p = \rho \frac{k}{\mu}\, 10^{b T} \ \ \ {(\mathrm{with}\ \mathrm{the}\ \mathrm{variables}\ p,\ \rho \ \mathrm{and}\ T)},
\end{equation}
or, equivalently, as
\begin{equation} \label{eq: EoS_PRhoT_expon}
     p = \rho \frac{k}{\mu}\, \mathrm{e}^{c T}.
\end{equation}
Here, $b$, $k$ and $c=b \ln 10$ are the positive constants. The coefficient $k$ has dimension of (energy)/(amount of substance), whereas dimension of the coefficients $b$ and $c$ is (temperature)$^{-1}$; and, as found, $k=4.1 \cdot 10^{3}$~J/mol, $b=28 \cdot 10^{-5}$~K$^{-1}$ and $c=64.47 \cdot 10^{-5}$~K$^{-1}$.

It is necessary to note some features of the found EoS that are relevant to the issue under consideration.

\noindent (i) In Eq.~(\ref{eq: EoS_PVT}), the system is specified only by the atomic mass $\mu$. The coefficients $k$ and $b$ (or $c$) are constant independent of the system. It is noteworthy that the coefficient $k$ is approximately equal to $(R/2) \cdot 10^3$~K.

\noindent (ii) Eq.~(\ref{eq: EoS_PVT}) satisfies empirical findings~(\ref{eq: first_property}), (\ref{eq: second_property}), (\ref{eq: third_property}) and represents the extensions of the perfect gas equation of state. According to this EoS, the energy term $PV$ increases nonlinearly with the temperature $T$ and much faster than for a perfect gas.

\noindent (iii) For the case of an alkali vapor, there is no transition to the ``stable'' thermodynamic regime of a perfect gas, which would appear in a some $(p,T)$-range. Nevertheless, there are isotherms $T_{id}$, on which the ideal gas law holds, and Eq.~(\ref{eq: EoS_PVT}) gives results that are identical to the results of equation $pV=(m/\mu)RT$. These isotherms can be found by calculating the compressibility factor $Z=(\mu pV/m RT)$, which is a dimensionless parameter  that estimates the deviation of the thermodynamic properties of the gas under consideration from the properties of a perfect gas~\cite{Thermodynamics,Kulinskiia_JCP,Reynolds_JCP}. From Eq.~(\ref{eq: EoS_PVT}), we find
\begin{equation} \label{eq: compressibility_factor}
    Z = \frac{k \; \mathrm{e}^{cT}}{RT}.
\end{equation}
Since for a perfect gas the compressibility factor is $Z=1$, then solving equation
\begin{equation}
    \frac{k \; \mathrm{e}^{cT}}{RT} = 1,
\end{equation}
we find the isotherms $T_{id}=856$~K and $2\,547$~K.

\noindent (iv) The results of Eq.~(\ref{eq: EoS_PVT}) should be well reproduced by equation of the form
\begin{equation} \label{eq: approx_EoS}
     p = \rho \frac{k}{\mu}\left ( 1 + cT + \frac{c^2 T^2}{2} + \frac{c^3 T^3}{6} + \ldots \right ),
\end{equation}
which follows directly from Eq.~(\ref{eq: EoS_PRhoT_expon}) by the Taylor series expansion of the exponential term over the dimensionless parameter $\xi=cT$. Taking into account that $c=64.47 \cdot 10^{-5}$~K$^{-1}$, one can estimate that the dimensionless parameter is $\xi \simeq 1$ when the temperatures are about $T =1\,550$~K.

\noindent (v) Similar to the perfect gas equation of state, Eq.~(\ref{eq: EoS_PVT}) does not predict any phase transitions or critical point. This equation is universal for all alkali elements and is intended solely to reproduce their thermodynamics in the equilibrium vapor phase. As follows from this equation, the thermodynamic principle of the corresponding state is fulfilled for the group of alkali elements. Unlike well-known thermodynamic gas models, such as the van der Waals model and its modified versions, the Redlich-Kwong model, the Peng-Robinson model, the Soave model, and others~\cite{Landau/Lifshitz_Stat,Skripov}, this principle is implemented here without reference to the critical parameters $p_c$, $\rho_c$ and $T_c$.

\begin{figure}[h!]
\begin{center}
    \includegraphics[keepaspectratio,width=0.52\linewidth]{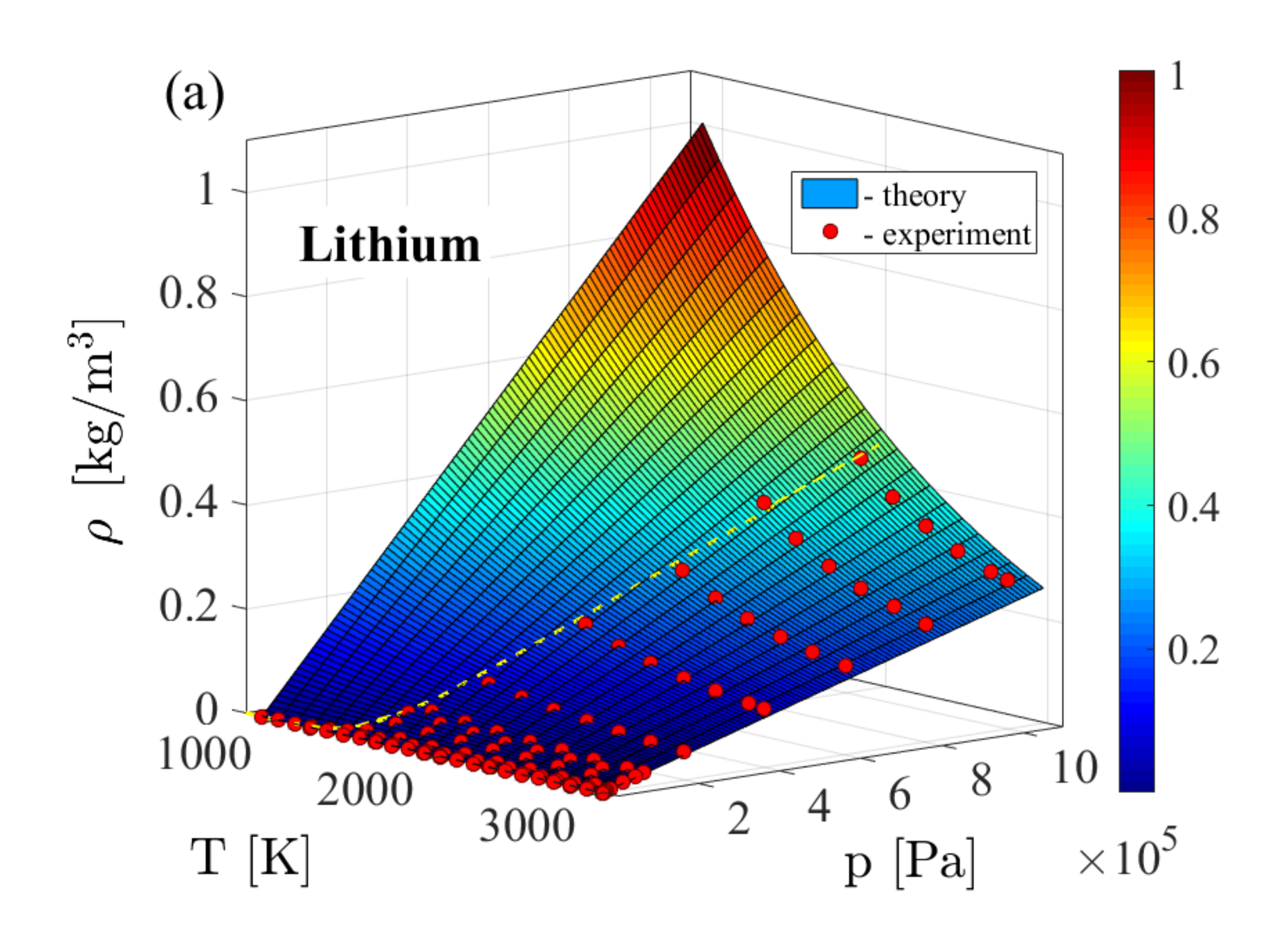} 
    \includegraphics[keepaspectratio,width=0.52\linewidth]{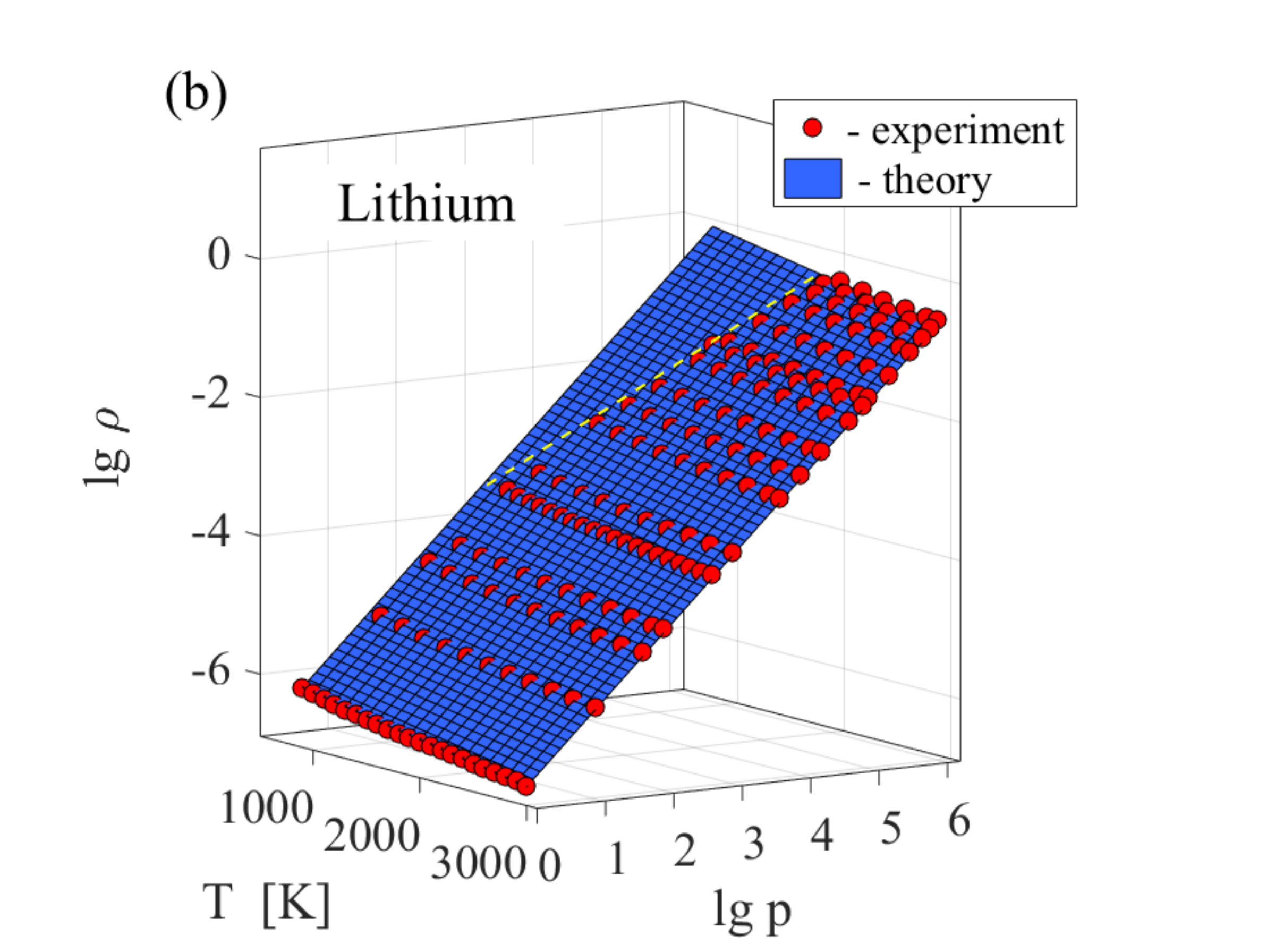}
    \includegraphics[keepaspectratio,width=0.52\linewidth]{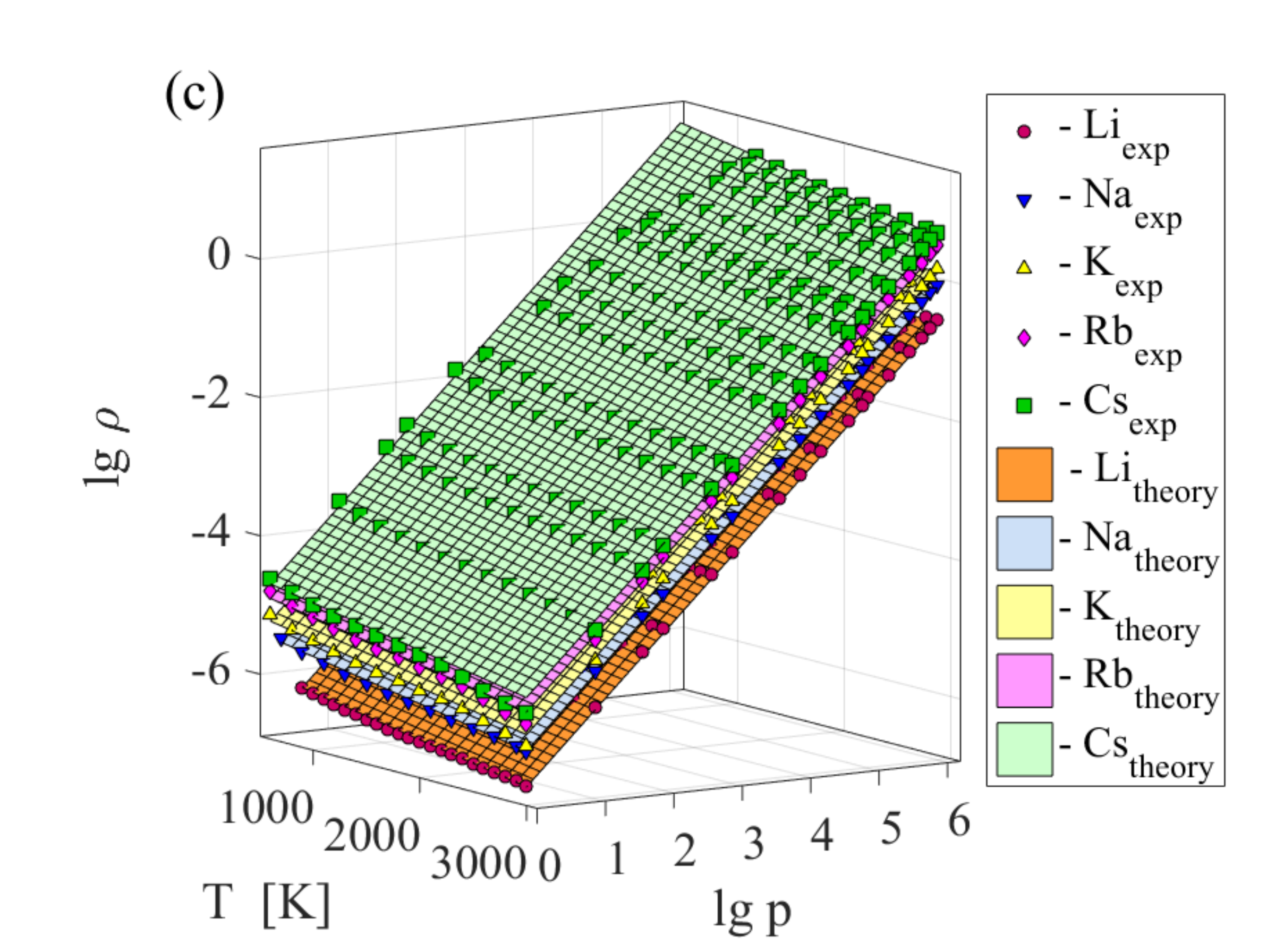}
\end{center}
\caption{Mass density as function of pressure and temperature of alkali elements in vapor phase. Symbols are experimental data~\cite{Vargaftik}, colored surfaces are theoretical results obtained using Eq.~(\ref{eq: EoS_PVT}); yellow dashed curves are the saturation curve evaluated by means of the Browning-Potter equation $\lg p_s = A - (B/T_s) - C \lg T_s$ with the parameters $A$, $B$ and $C$ taken from Ref.~\cite{Thermophys_prop}. (a) Thermal EoS $f(p,\rho,T)=0$ of lithium vapor. (b) Thermal EoS $f(\lg p,\lg \rho,T)=0$ of lithium vapor. (c) Thermal EoS $f(\lg p,\lg \rho,T)=0$ of various alkali elements in vapor phase.}
\label{fig: EoS_Li}
\end{figure}
\begin{figure}[h!]
\begin{center}
    \includegraphics[keepaspectratio,width=0.54\linewidth]{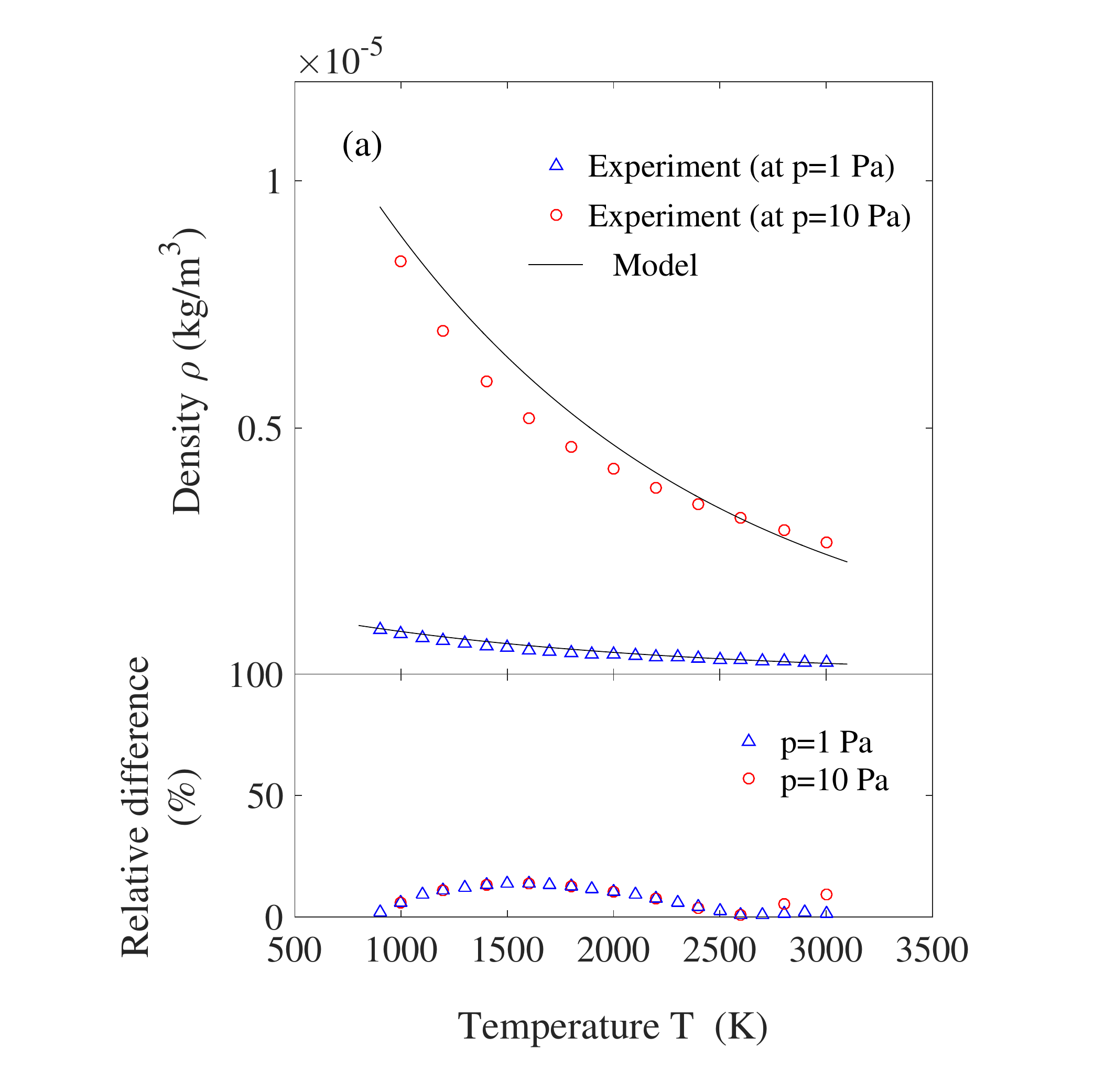} \hskip -1.1 cm
    \includegraphics[keepaspectratio,width=0.54\linewidth]{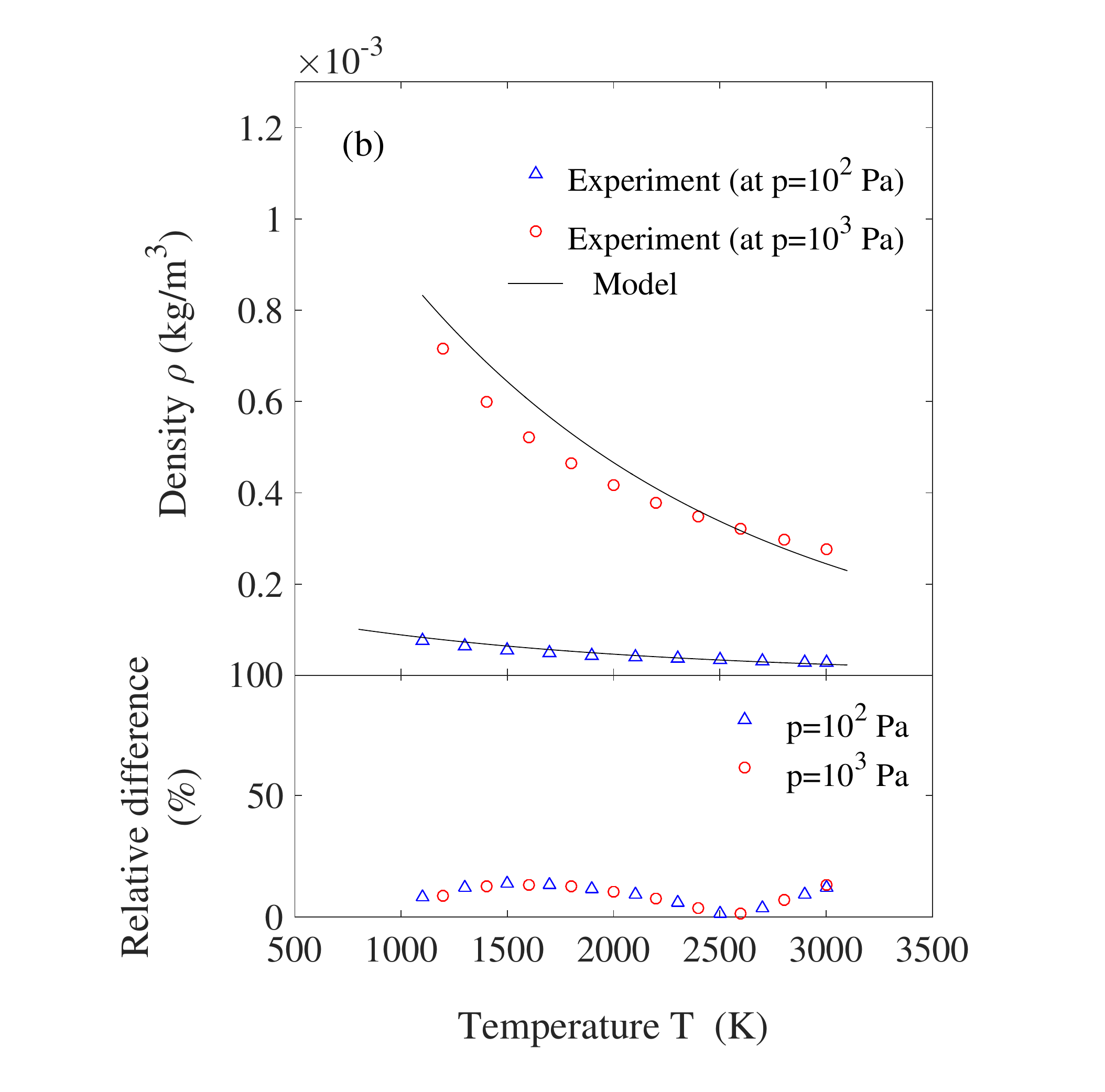}
    \includegraphics[keepaspectratio,width=0.54\linewidth]{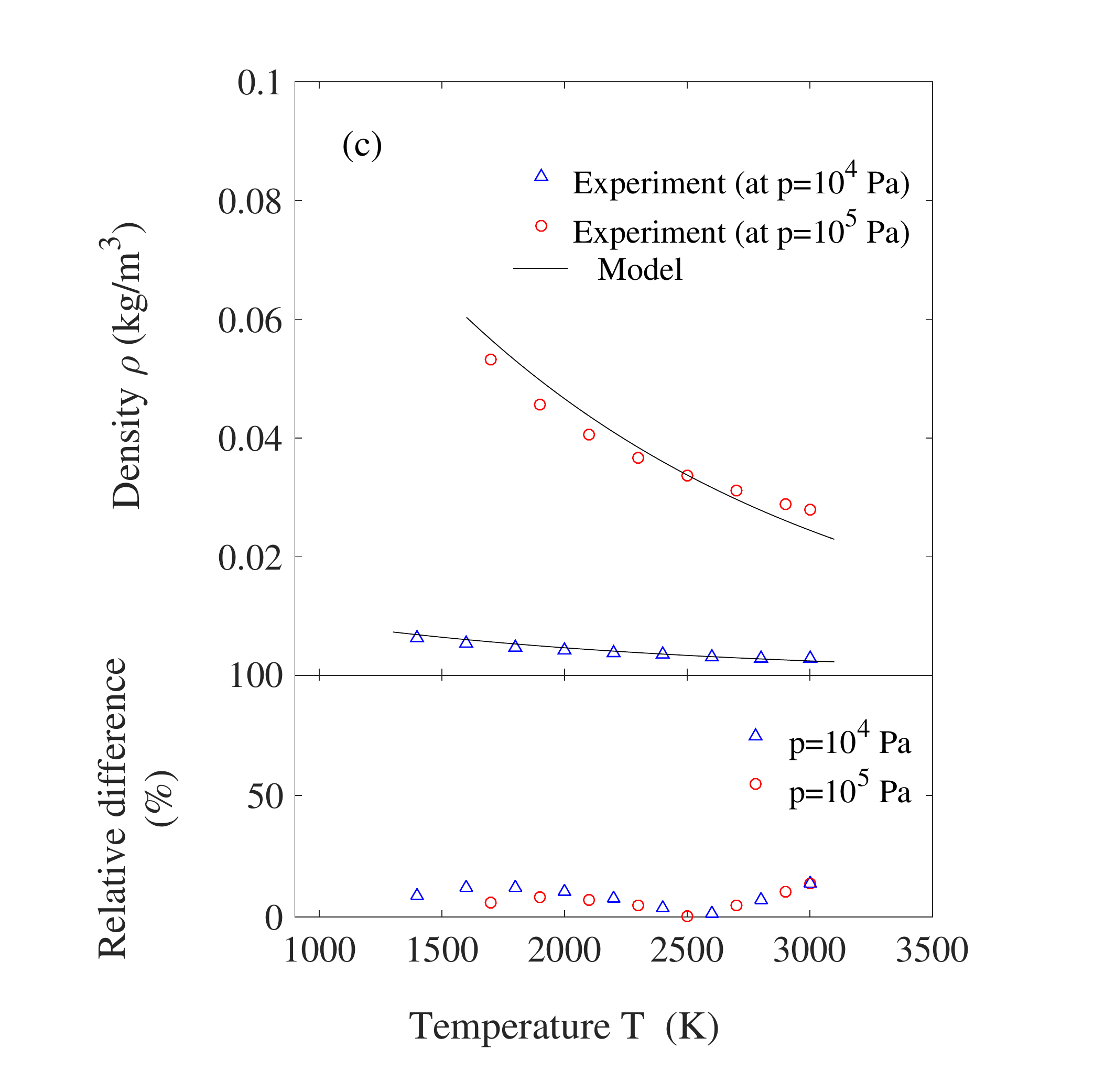}
\end{center}
\caption{Mass density of lithium vapor as function of temperature at fixed values of pressure (top parts in figures) and relative difference between experimental data and theoretical results with Eq.~(\ref{eq: EoS_PVT}) (lower parts in figures). (a) Results for isobars $p=1$ and $10$~Pa; (b) for isobars $p=10^2$ and $10^3$~Pa; and (c) for isobars $p=10^4$ and $10^5$~Pa.
}
\label{fig: 2D_EoS_Li}
\end{figure}
In Figs.~\ref{fig: EoS_Li}(a) and \ref{fig: EoS_Li}(b), as an example, the results of Eq.~(\ref{eq: EoS_PRhoT}) are compared with the experimental data for the case of lithium vapor~\cite{Vargaftik}. As seen, Eq.~(\ref{eq: EoS_PRhoT}) correctly reproduces the density landscape as a function of the pressure $p$ and the temperature $T$ over the entire ($p, T$)-range corresponding to equilibrium vapor phase. Further, in both Figs.~\ref{fig: EoS_Li}(a) and \ref{fig: EoS_Li}(b), the supersaturated vapor density $\rho$ predicted by Eq.~(\ref{eq: EoS_PRhoT}) for the ($p, T$)-range above the saturation curve $p_s=p_s(T_s)$ is also shown.
Full agreement of theoretical results with experimental data is observed for other alkali elements as well [see Fig.~\ref{fig: EoS_Li}(c)]. In the double logarithmic representation, the data for EoS for various systems,  $f(\lg p,\lg\rho,T)=0$, form parallel planes, spaced with respect to the variable $\lg \rho$. These features are properly reproduced by Eq.~(\ref{eq: EoS_PRhoT}) that can be rewritten in the form of equation of a plane:
\begin{equation} \label{eq: EoS_plane}
    \lg p - b T - \lg \rho - \lg \frac{k}{\mu} = 0
\end{equation}
with the found values of the coefficients $b$ and $k$ (see video in Supplemental material).

The isobars $\rho = \rho(T)$ for lithium vapor are shown in Fig.~\ref{fig: 2D_EoS_Li}, where the theoretical results from Eq.~(\ref{eq: EoS_PVT}) are compared with experimental data. Note that the results in Fig.~\ref{fig: 2D_EoS_Li} are completely identical to those shown in Figs.~\ref{fig: EoS_Li}(a) and \ref{fig: EoS_Li}(b); and difference is only in the representation of these results. Here, in each panel of the figure, the results are given for a pair of isobars. The theoretical model reveals good agreement with experimental data over the entire wide pressure range, where the mass density $\rho$ changes significantly and, namely, by four orders of magnitude. The relative difference between the theoretical results and experimental data practically does not exceed $10$~$\%$. It is noteworthy that the model properly reproduces the temperature dependences of the mass density, the changes in these dependences with pressure, as well as the density values near saturation.

\subsection{The thermal coefficients}
The analytical EoS allows one to evaluate the thermal coefficients~\cite{Landau/Lifshitz_Stat}: the volumetric coefficient of thermal expansion
\begin{equation} \label{eq: thermal_expan}
    \alpha = -\frac{1}{\rho} \left( \frac{\partial \rho}{\partial T} \right )_p,
\end{equation}
the thermal pressure coefficient
\begin{equation} \label{eq: pressure_coef}
    \beta = \frac{1}{p} \left( \frac{\partial p}{\partial T} \right )_\rho
\end{equation}
and the isothermal compressibility coefficient
\begin{equation} \label{eq: isotherm_compressibility}
    \kappa_T = \frac{1}{\rho} \left ( \frac{\partial \rho}{\partial p} \right )_T .
\end{equation}
From Eqs.~(\ref{eq: EoS_PRhoT}), (\ref{eq: thermal_expan}), (\ref{eq: pressure_coef}) and (\ref{eq: isotherm_compressibility}), we obtain the results typical of a perfect gas scenario:
\begin{equation}
    \kappa_T = \frac{1}{p}
\end{equation}
and
\begin{equation} \label{eq: thermal_expan_our}
    \alpha = \beta.
\end{equation}
On the other hand, the thermal expansion coefficient~$\alpha$  and the pressure coefficient~$\beta$ in Eq.~(\ref{eq: thermal_expan_our}) are not functions of temperature, as in the case of a perfect gas, and are defined as
\begin{equation}
    \alpha = \beta = c.
\end{equation}
Thus, we find that both the coefficients, $\alpha$  and $\beta$, take the value $\sim 64.5 \cdot 10^{-5}$~K$^{-1}$ for all alkali elements.

\subsection{Caloric equation of states and the internal energy}
The caloric equation of state shows how the internal energy $U$ is related to the volume $V$ and the temperature $T$ of considered system, i.e. $U=U(V,T)$, or, in general, with any pair of the parameters from the following: $p$, $V$ and $T$. Taking into account the thermodynamic identity
\begin{equation} \label{eq: thermodyn_identity}
    T \left (  \frac{\partial p}{\partial T} \right )_V =  \left (  \frac{\partial U}{\partial V} \right )_T + p,
\end{equation}
we find from Eq.~(\ref{eq: EoS_PVT})  the caloric EoS
\begin{equation}
\label{eq: caloric_EoS}
      \left( \frac{\partial U}{\partial V} \right)_T = \frac{m}{V}  \frac{k}{\mu} \mathrm{e}^{cT} (c T  - 1).
\end{equation}
As follows from this, the internal energy $U$ depends only on the temperature $T$, that again is similar to the perfect gas scenario with the $T$-dependent internal energy, i.e. $U_{id}=U_{id}(T)$. Nevertheless, the character of the $T$-dependent internal energy of alkali vapors will differ from the case of a perfect gas, where $U_{id}(T)=(3/2)(m/\mu)RT$.  From Eq.~(\ref{eq: caloric_EoS}), we get that the specific internal energy $u = (\mu/m) U$ can be written as
\begin{equation} \label{eq: specific_int_energy}
    u(T)  = \frac{3}{2}RT + k\,\mathrm{e}^{cT}\left ( cT-1 \right).
\end{equation}
Thus, for the temperatures corresponding to the vapor phase of alkali elements, the quantity $u(T)$ increases with temperature faster in comparison with a simple linear $T$-dependence: $u_{id}(T) \propto T$. This is due to the ionization processes in an alkali high temperature vapor. The number of the charged components  including free electrons and ions of the various sizes increase with temperature in the system~\cite{Fortov_Extereme,Physica_4,Likalter_PRB_1996,Likalter_2002,Hensel_1999,Mokshin_PRL_2004}. In addition, the interactions between these charged components have a long-range character, and, actually, the perfect gas regime turns out to be unattainable.

\subsection{The heat capacities}
The molar heat capacities, $c_V$ and $c_p$, are defined by well known thermodynamic relations
\begin{equation} \label{eq: C_V}
    c_V = \frac{\mu}{m} \left ( \frac{\partial U}{\partial T} \right )_V
\end{equation}
and
\begin{equation} \label{eq: C_P}
    c_p = c_V + \frac{\mu}{m} \frac{\alpha^2}{\kappa_T} VT.
\end{equation}
Then, taking into account Eq.~(\ref{eq: caloric_EoS}), we obtain the molar heat capacities
\begin{equation} \label{eq: c_V_alk}
    c_V = \frac{3}{2}R + k c^2 T \mathrm{e}^{cT}
\end{equation}
and
\begin{eqnarray} \label{eq: c_P_alk}
    c_p &=& \frac{3}{2}R + 2 k c^2 T \mathrm{e}^{cT} \nonumber \\
        &=& 2 c_V - \frac{3}{2}R.
\end{eqnarray}
As expected, both the heat capacities $c_V$ and $c_p$ defined by Eqs.~(\ref{eq: c_V_alk}) and (\ref{eq: c_P_alk}) depend only on temperature and increase with temperature in a similar way. Over the considered temperature range $T \in [600;\; 3\,000]$~K, the quantity $c_V$ exceeds the value $(3/2)R$ corresponding to a monatomic perfect gas; it increases threefold over this temperature range. The quantity $c_p$ exceeds the perfect gas value $(5/2)R$ starting from the temperature $T=1\,157$~K. It is noteworthy that the difference in the molar heat capacities $(c_p - c_V)$ and the heat capacity ratio $\gamma = c_p / c_V$ should also demonstrate an increase with temperature, as follows from expressions
\begin{equation}
\label{eq: hc_difference}
      c_p - c_V =  k c^2 T \mathrm{e}^{cT}
\end{equation}
and
\begin{equation}
\label{eq: gamma}
      \gamma = 2 - \frac{3}{2}\frac{R}{c_V}.
\end{equation}
Near saturation the heat capacity ratio $\gamma$ takes a value comparable to unity, that is typical for liquid alkali metals~\cite{Ohse_Handbook}, and it grows with temperature~\cite{Fink_Sodium}.
In particular, according to Eq.~(\ref{eq: gamma}), the parameter $\gamma$ is about $1.11$  at the temperature $T=600$~K and it is about $1.74$ at $T=3\,000$~K.

\section{Conclusion}

In conclusion, we note the following points. This study shows that for the considered ($p,T$)-range, the thermodynamics of alkali vapors can be evaluated using a fairly simple model equation of state, which is able to correctly capture the main features of equilibrium vapor thermodynamics. Due to its simple analytical form, some fine thermodynamic effects are ignored. For example, for the same temperature range this description can provide the same quantitative results for all the alkali systems. A more accurate theoretical description can be performed using more rigorous theoretical approaches such as the droplet model of a nonideal alkali plasma (see the studies of Fortov, Ebeling, R\"opke and coworkers) or the moment theory developed within the Bogoliubov's ideas~\cite{Fortov_plasma,Sadykova,Mokshin_TMF_2015}. The EoS considered in the given study is of the type of single-phase equation of states; it does not predict any phase transitions and critical point parameters, nevertheless, it allows one to evaluate the thermodynamic properties of supersaturated vapors. Further, although there are some similarities with the perfect gas equation of state, the given EoS for the considered ($p,T$)-range of alkali vapors does not reveal the transition into the regime of a perfect gas. The results of this study confirm that the thermodynamic principle of the corresponding states is applicable to describe the thermodynamic properties of alkali metals in the equilibrium gas phase.

\vskip 0.3 cm

\section*{CRediT authorship contribution statement}
\textbf{Anatolii V. Mokshin}: Supervision, Conceptualization, Investigation, Writing –- review \& editing. \textbf{Diana A. Mirziyarova}: Data analysis, Investigation, Writing –- original draft.

\vskip 0.3 cm

\section*{Declaration of competing interest}
The authors declare that they have no known competing financial interests or personal relationships that could have appeared to influence the work reported in this paper.

\vskip 0.3 cm

\section*{Acknowledgments}
This work was supported by the Russian Science Foundation (Project No. 19-12-00022). The authors are grateful to Bulat Galimzyanov and Roman Khabibullin for discussions.

\end{document}